\documentstyle[twocolumn,aps,prb,psfig]{revtex}
\begin{document}
\title{Coherent Control for a Two-level System Coupled to Phonons}
\author{H. Castella}
\address{Max-Planck-Institut f\"ur Physik komplexer Systeme,\\ 
N\"othnitzer Str. 38, D-01187 Dresden, Germany.}
\author{R. Zimmermann}
\address{Institut f\"ur Physik der Humboldt-Universit\"at zu Berlin,\\ 
Hausvogteiplatz 5-7, D-10117 Berlin, Germany.}
\date{Received \today }
\bigskip\bigskip
\maketitle
\begin{abstract}
The interband polarizations induced by two phase-locked pulses in a
semiconductor show strong interference effects depending on the time 
$\tau_1$ separating the pulses. The four-wave mixing signal diffracted from a 
third pulse delayed by $\tau$ is coherently controlled by tuning 
$\tau_1$. The four-wave mixing response is evaluated exactly for a two-level 
system coupled to a single LO phonon. In the weak coupling regime it 
shows oscillations with the phonon frequency which turn into sharp peaks 
at multiples of the phonon period for a larger coupling strength. Destructive 
interferences between the two phase-locked pulses produce a splitting of the 
phonon peaks into a doublet. For fixed $\tau$ but varying $\tau_1$ the signal 
shows rapid oscillations at the interband-transition frequency, whose 
amplitude exhibits bursts at multiples of the phonon period.
\end{abstract}
\pacs{PACS numbers:78.47.+p, 42.50.Md, 42.65.Re, 63.20.Kr}
Ultrafast spectroscopy on semiconductors gives a direct insight into
the loss of coherence between optically excited carriers. The dephasing
takes place through different mechanisms such as emission of 
phonons, Coulomb interaction or scattering by impurities. In the first hundred 
femtoseconds after an optical excitation few scattering processes have 
occurred and the electronic wavefunction may be described by a
superposition of different states with well defined phase relations.
The resulting quantum beats are a signature of phase coherence. 

Quantum beats with the longitudinal optical (LO) phonon frequency have been 
observed in four-wave mixing (FWM) experiments on GaAs which was excited 
resonantly at the excitonic energy \cite{weg1}. The beatings are due to 
interference between states with different number of virtual phonons.
Indeed electrons close to the band edge which do not have enough energy to 
emit any real LO phonon, form a polaronic state. 

The phase coherence of the excitation may also be used to coherently control
the optical response of the sample via a pair of phase-locked 
pulses which induce interfering polarizations in the sample \cite{bam}. 
Recently this coherent control was applied to the quantum beats in the
FWM signal \cite{weg2}. Depending on the time delay $\tau_1$ between 
phase-locked pulses the amplitude of the quantum beats was modulated,
thus experimentally demonstrating that the phase coherence is not lost in the 
first few scattering processes.

The present work computes the FWM response in third order in the field for a 
simple two-level system treating exactly the coupling to a single phonon mode. 
The calculations closely follow Ref. \onlinecite{mah} which studied the linear 
response in connection to phonon broadening of impurity spectra in 
semiconductors \cite{mah2}. Recently the FWM response in the same model was 
obtained for $\delta$-function pulses by solving the infinite hierarchy of 
kinetic equations \cite{axt}.

We focus here on the regime of strong coupling between electrons and
phonons and discuss the recent coherent-control experiment on the polar 
material ZnSe, where peaks separated by approximately 
half the phonon period were observed and interpreted as signatures of
two-phonon processes \cite{weg3}. In the two-level system, however, we 
attribute these features to a splitting of the phonon peaks into
doublets due to destructive interference between the signals generated
by the two phase-locked pulses. We further predict 
non-perturbative bursts of the FWM signal as a function of delay between
phase-locked pulses at multiples of the phonon period $T_{ph}$.

A two-level system $\mid i\rangle$, $i=0,1$ with energies 
$\epsilon_0<\epsilon_1$ couples to the
position $a+a^{\dagger}$ of a harmonic oscillator of frequency $\Omega$
with a level-dependent coupling strength $g_i$. In addition the classical 
light field $E(t)$ induces transitions between the two levels:
\begin{eqnarray}
H=&\sum_{i=0,1}\left(\epsilon_i+g_i(a+a^{\dagger})\right)
\mid i\rangle\langle i\mid +\Omega a^{\dagger}a
\nonumber\\
&-\mu E(t)
\bigl(\mid 0\rangle\langle 1\mid +H.c.\bigr).
\label{eq1}
\end{eqnarray}

This model describes the coupling of electrons to LO phonons
of frequency $\Omega=2\pi/T_{ph}$ using a single oscillator. The two levels may 
be viewed as states in the valence and conduction bands which are resonantly 
excited by light. Within this
interpretation the coupling constants $g_0$ and $g_1$ should be taken equal 
since the Fr\"ohlich interaction does not depend on the effective mass of the 
electrons i.e. on the band \cite{rid}.
Another picture associates the level $\mid 0\rangle$ to the full valence 
and empty conduction band while $\mid 1\rangle$ corresponds to an excitonic 
state which has an effective coupling \cite{seg} $g_1$ to the phonons 
differing from the ground-state value $g_0$. This interpretation is 
particularly meaningful for the FWM experiments where the
excitation is resonant with the excitonic level. 

The model considers only scattering of the electron within the same level and
hence cannot describe relaxation effects. This approximation which neglects the
recoil of the electron is particularly relevant to the strong coupling 
regime since it corresponds to the small-polaron limit\cite{mah,sma}. 

\begin{figure}[htb]
\centerline{\psfig{file=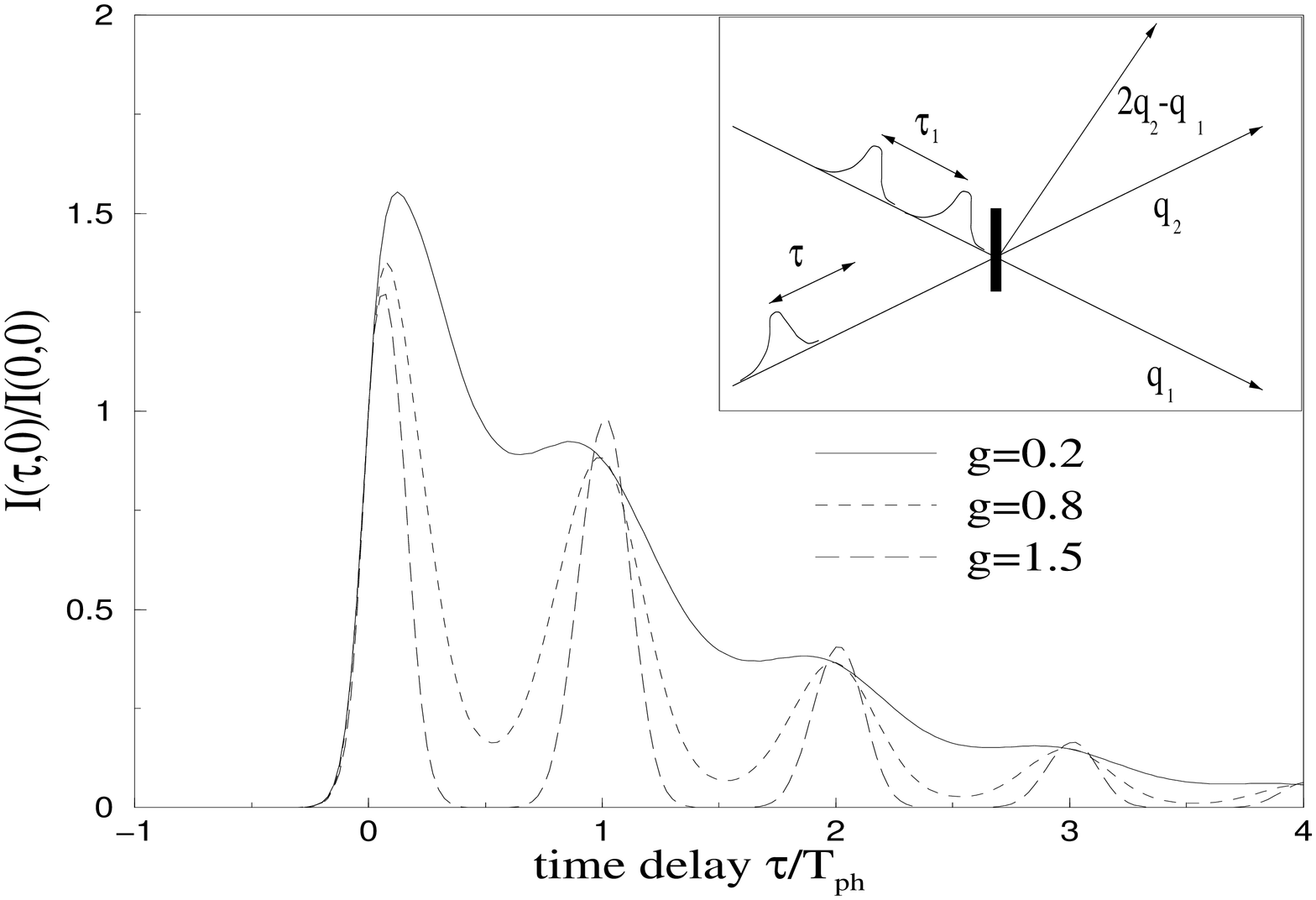,width=3.375in,height=3in,angle=-90}}
\caption{Normalized integrated FWM signal $I(\tau,\tau_1=0)/I(0,0)$ as a 
function of delay $\tau$ in units of the phonon period $T_{ph}$ for 
a single pulse in direction $q_1$, i.e. $\tau_1=0$.
The weak modulation of the signal for $g=0.2$ turn into distinct peaks at 
multiples of the phonon period with increasing coupling. The inset shows the 
setup for the coherent control where two phase-locked pulses delayed by 
$\tau_1$ propagate in direction $q_1$. A third pulse in direction $q_2$ 
delayed by $\tau$ is diffracted along $2q_2-q_1$.}
\label{f1}
\centerline{\psfig{file=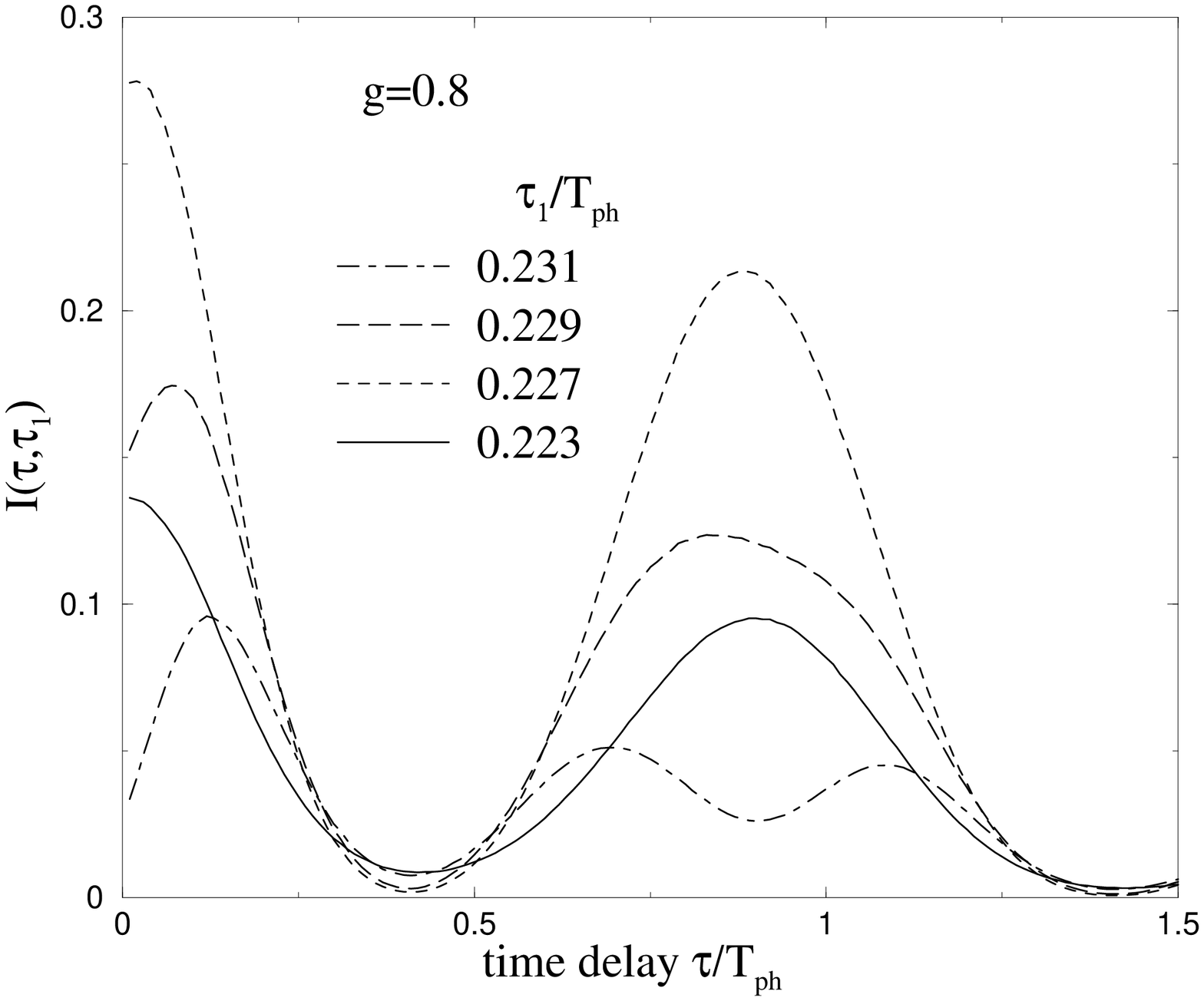,width=3.375in,height=3.in,angle=-90}}
\caption{Integrated FWM signal $I(\tau,\tau_1)$ as a function of delay 
$\tau$ for different $\tau_1$ around $\epsilon\tau_1\simeq 41\pi$. 
Constructive interferences between phase-locked pulses produce a 
single peak at $\tau=T_{ph}-\tau_1/2$ which splits into a doublet when
$\tau_1$ is tuned to achieve destructive interferences.}
\label{f2}
\centerline{\psfig{file=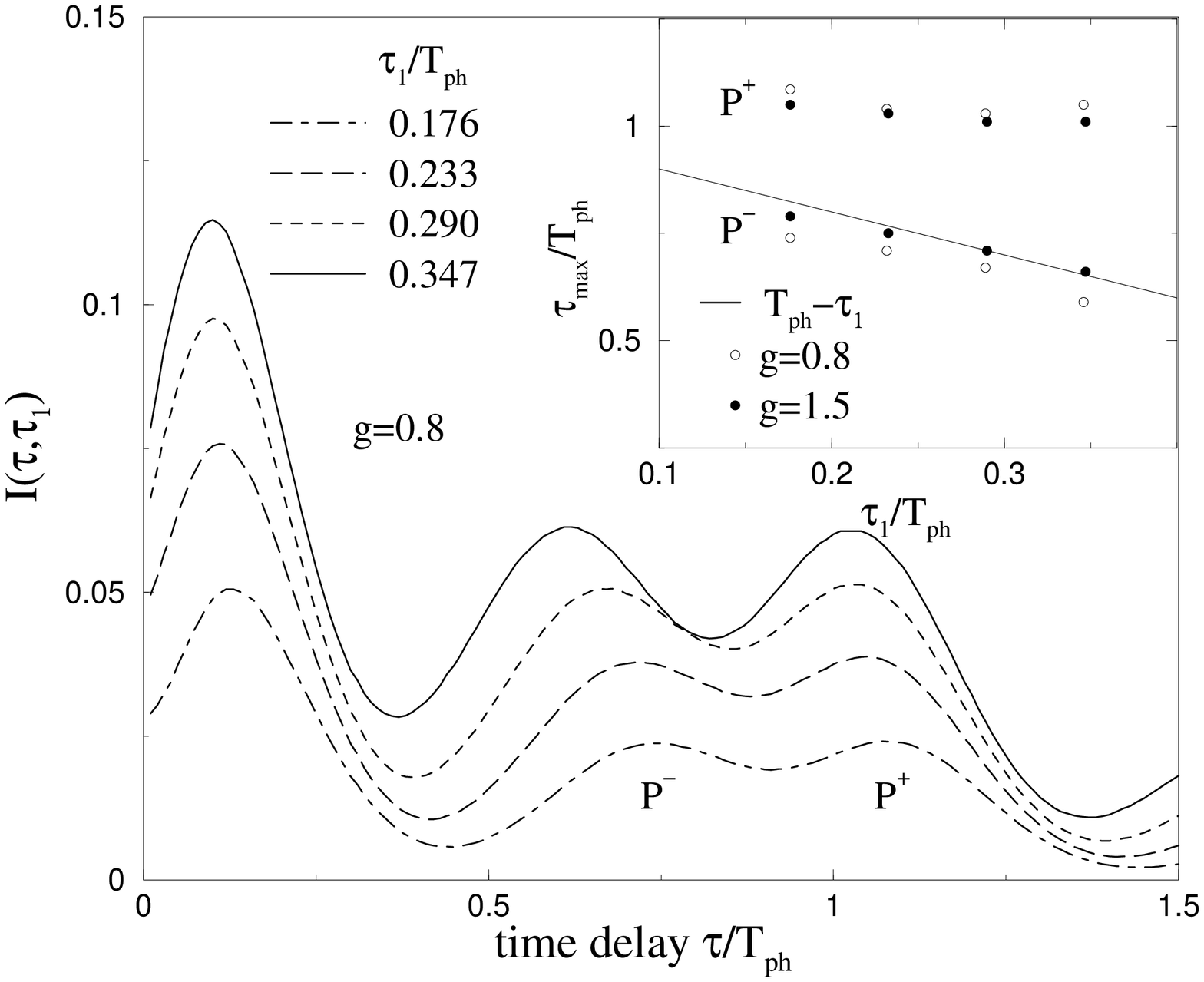,width=3.375in,height=3.in,angle=-90}}
\caption{Integrated FWM signal $I(\tau,\tau_1)$ for $g=0.8$ as a function of 
time delay $\tau$ for different times $\tau_1$. The inset shows the position 
of the maxima $\tau_{max}$ of the two peaks $P^-$ and $P^+$ as a function of 
$\tau_1$ for $g=0.8$ and $1.5$. With increasing coupling, the dispersion of 
the $P^-$ peak approaches $T_{ph}-\tau_1$ plotted as the solid line.}
\label{f3}
\centerline{\psfig{file=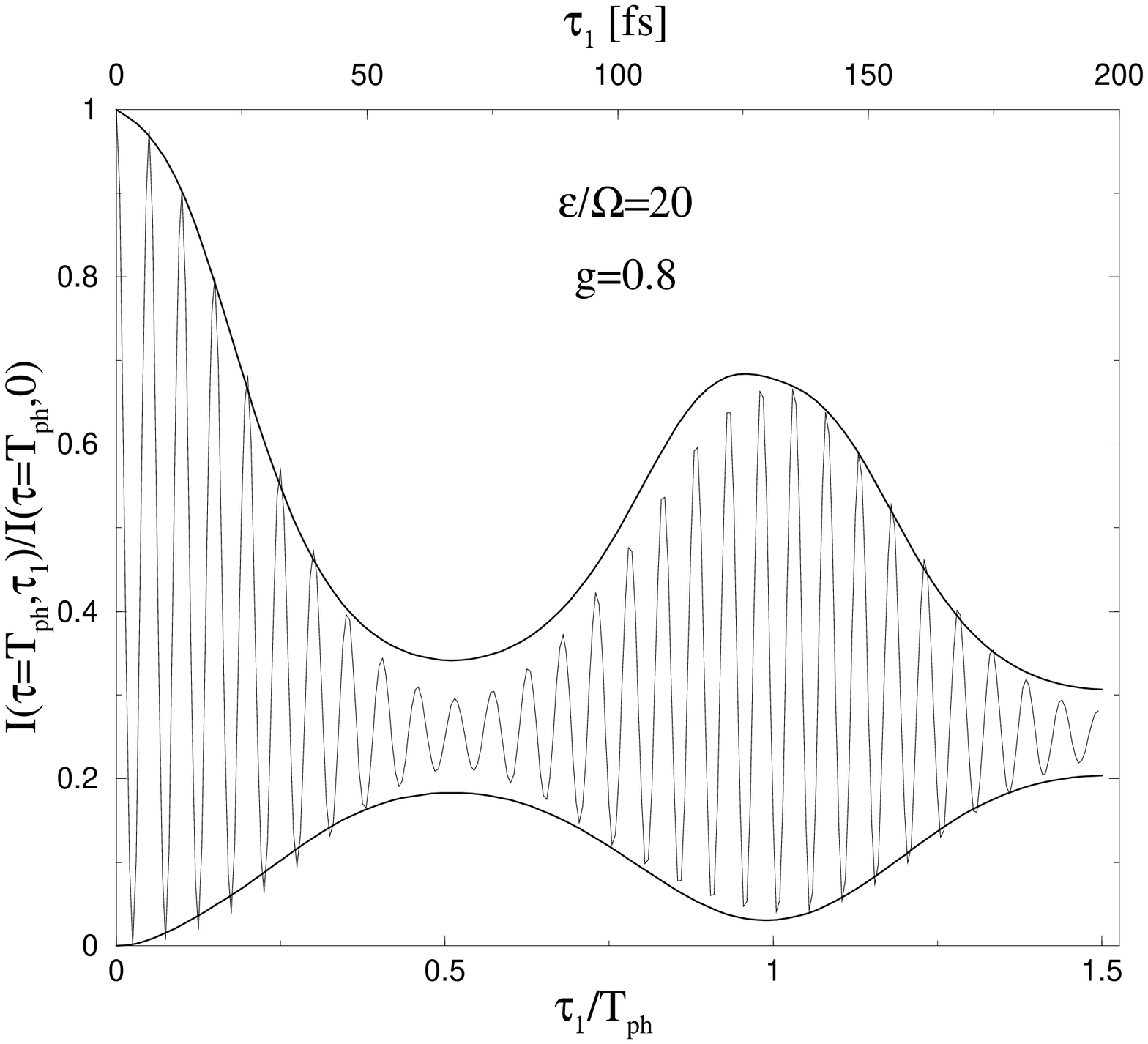,width=3.375in,height=3.in,angle=-90}}
\caption{Normalized intensity of the first phonon peak 
$I(\tau=T_{ph},\tau_1)/I(\tau=T_{ph},0)$ as a function of delay time 
$\tau_1$ between phase-locked pulses. The rapid oscillations of
frequency $\epsilon$ which are first suppressed with increasing $\tau_1$, 
show bursts at later times when $\tau_1$ is a multiple of the phonon period 
$T_{ph}$. The envelopes of the oscillations are well 
described by Eq.(8) valid for large $g$ (thick lines). For better display
of the interference patterns, the frequency $\epsilon$ has been reduced by
a factor of 4 from its value for ZnSe.}
\label{f4}
\end{figure}

{\it Four-wave mixing}. A first pulse $E_1(t)$ creates an interband
polarization in the sample while a second pulse $E_2(t-\tau)$ delayed by 
a time $\tau$ is diffracted along the direction
$2q_2-q_1$ where $q_1$ and $q_2$ are the propagation
directions for the first and second pulses, respectively (see inset of
Fig.\ref{f1}). The FWM signal $F(t,\tau)$ at time $t$ and delay $\tau$ is 
related to the third order response \cite{sha} $\chi^{(3)}$: 
\begin{eqnarray}
F(t,\tau)=-i\mu^3\int_{-\infty}^{\infty}dt_1dt_2dt_3 E^{\ast}_1(t_2)
\nonumber\\
E_2(t_1-\tau)
E_2(t_3-\tau)
\chi^{(3)}(t-t_1,t-t_2,t-t_3)
\label{eq2}
\end{eqnarray}

The electron occupies the lowest level $\mid 0\rangle$ before the pulses 
arrive, and the oscillator is in thermal equilibrium at an inverse temperature 
$\beta=1/k_BT$. 
The pulses cause transitions between electronic levels which affect the 
evolution of the oscillator via the level dependent Hamiltonian 
$H_i=\Omega a^{\dagger}a+\epsilon_i+g_i(a+a^{\dagger})$.
The response function involves a thermal average of evolution operators 
and has a main contribution at positive time delays and 
a second term for $t_1>t_2>t_3$ i.e. when the two pulses overlap:
\begin{eqnarray}
\chi^{(3)}(t_1,t_2,t_3)=
\Theta(t_1)\Theta(t_2-t_1)\times
\nonumber\\
\Big(\Theta(t_3)\langle e^{iH_1(t_2-t_1)}e^{iH_0t_1}
e^{-iH_1t_3} e^{iH_0(t_3-t_2)}\rangle+
\nonumber\\
\Theta(t_3-t_2)\langle e^{-iH_1t_1}e^{iH_0(t_2-t_1)}
e^{iH_1(t_2-t_3)} e^{iH_0t_3}\rangle\Big).
\label{eq3}
\end{eqnarray}

The evaluation of the terms in brackets follows exactly 
Ref.\onlinecite{mah}. The Hamiltonian $H_1$ is diagonalized by the 
Lang-Firsov transformation \cite{lan} $S=g(a^{\dagger}-a)$:
$\exp(S)H_1\exp(-S)=H_0+\epsilon$ where
$\epsilon=\epsilon_1-\epsilon_0+(g_0^2-g_1^2)/\Omega$ is
the renormalized transition energy and $g=(g_1-g_0)/\Omega$ 
a dimensionless coupling constant. Since only the difference 
$g_1-g_0$ enters the transformation, it is essential that
e.g. excitonic effects renormalize the coupling strengths
from the non-interacting value $g_1=g_0$.

The product of evolution operators in Eq.(\ref{eq3}) is evaluated using the 
previous canonical transformation and the following relation between the 
time-dependent operators $S(t)=\exp(iH_0t)S\exp(-iH_0t)$:
\begin{equation}
e^{S(t_1)}e^{-S(t_2)}=e^{S(t_1)-S(t_2)}e^{-ig^2\sin(\Omega(t_1-t_2))}.
\label{eq4}
\end{equation}

We do not reproduce here the lengthy expression for the response function
but instead discuss the FWM signal $F(t,\tau)$ for $\delta$-function
pulses. The signal oscillates at the renormalized energy $\epsilon$ with a
time delay $2\tau$ typical of a photon echo. The polaronic nature of the state 
shows up in the exponential dependence on the coupling constant, and the
temperature enters via the Bose function $N_B=1/(\exp(\beta\Omega)-1)$:
\begin{eqnarray}
F(t,\tau)=-i\Theta(\tau)\Theta(t-\tau)e^{-i\epsilon(t-2\tau)}
\times
\nonumber\\
e^{-g^2\bigl((N_B+1/2)\mid 2e^{i\Omega\tau}-e^{i\Omega t}-1\mid^2
-2i\sin(\Omega\tau)+i\sin(\Omega t)\bigr)}.
\label{eq5}
\end{eqnarray}
The FWM signal which depends simply on $\Omega\tau$ exhibits quantum beats at 
harmonics of the bare LO frequency, in contrast to the frequency 
shift seen experimentally \cite{weg1}. While the intraband scattering is 
missing in the present exactly solvable model, the simulations of 
Ref.\onlinecite{weg1} for a full two-band model do reproduce the correct 
frequency within approximate quantum kinetic equations. 
Moreover the FWM signal in Eq.(\ref{eq5}) has evenly 
spaced phonon replica in the frequency domain, which are not observed in 
experiments.

{\it Coherent control}. Two phase-locked pulses of equal amplitude propagate 
in direction $q_1$ and are separated by a time delay\cite{com} $\tau_1$ as 
shown in the inset of Fig. \ref{f1}. Since the FWM is linear in the field 
$E_1(t)$, the total signal is simply $F(t,\tau)+F(t+\tau_1,\tau+\tau_1)$ which 
shows strong interference effects depending on $\tau_1$ through the relative 
phase of the induced polarizations.

The time-integrated FWM $I(\tau,\tau_1)=\int_{-\infty}^{\infty}\mid F(t,\tau)+
F(t+\tau_1,\tau+\tau_1)\mid^2dt$ can be calculated analytically
for $\delta$-function pulses, $E_1(t)=\delta(t)+\delta(t+\tau_1)$.
A phenomenological lifetime $\Gamma$ which accounts for the 
damping of the polarization is introduced as an imaginary part of
$\epsilon$. For simplicity we give here only the result at zero temperature 
($N_B=0$) and for $\Gamma\ll\Omega$, which involves the modified Bessel 
function $I_0$:
\begin{equation}
I(\tau,\tau_1)=G(\tau,0)+G(\tau+\tau_1,0)+ 2 G(\tau+\frac{\tau_1}{2},
\tau_1).
\label{eq6}
\end{equation}
\begin{eqnarray}
G(\tau,\tau_1)=\frac{e^{-2\Gamma\tau}}{2\Gamma}Re\Biggl[
e^{-i\epsilon\tau_1-2g^2(3-2e^{-i\Omega\tau_1/2}\cos(\Omega\tau))}\times
\nonumber\\
I_0\left(2g^2\left((2-e^{i\Omega(\tau-\tau_1/2)})
(2-e^{-i\Omega(\tau+\tau_1/2)})
\right)^{1/2}\right)\Biggr].
\end{eqnarray}

We first discuss the contribution of a single pulse when the integrated FWM
equals $G(\tau,0)$ which decays exponentially 
with time delay and is modulated by terms oscillating with multiples
of $\Omega$. In the weak coupling regime, only the first harmonics 
$\cos(\Omega\tau)$ contributes to the modulation as given explicitely in
Ref. \onlinecite{axt}. Figure \ref{f1} shows the integrated FWM response 
$I(\tau,\tau_1=0)$ calculated  with the exact $\chi^{(3)}$ for pulses of
finite width which are taken as sech$^2$-shaped pulses resonant with 
the two levels and of 13fs duration. The parameters \cite{weg3} for ZnSe are 
$\Omega=32$meV, $\epsilon=2.8$eV, $T=77$K and $1/\Gamma=300$fs. The weak 
modulation of the signal for a small coupling $g=0.2$ relevant to GaAs, evolves 
into distinct peaks at multiples of the phonon period for a larger coupling
$g=0.8$ as in ZnSe. In the strong coupling regime indeed, the many 
harmonics sum up into sharp peaks of width proportional to $1/\sqrt{g}$: 
$G(\tau,0)\simeq\exp(-\Gamma\tau-16g^2\sin^4(\Omega\tau/2))$.

Two phase-locked pulses in direction $q_1$ produce interference effects 
depending on the delay $\tau_1$. When the two levels are decoupled from 
the phonons ($g=0$), the FWM signal in Eq. (\ref{eq6}) oscillates as 
$\sin^2(\epsilon\tau_1/2)$, and is thus suppressed when the induced 
polarizations are exactly out of phase, i.e. $\epsilon\tau_1=(2n+1)\pi$.

In the strong coupling regime, the first two terms on the 
right-hand side of Eq. (\ref{eq6}) correspond to the separate responses
of each pulse which are peaked at $\tau=nT_{ph}$ and $nT_{ph}-\tau_1$, 
respectively. The third term describes the interference between polarizations 
and has its maximum at $\tau=nT_{ph}-\tau_1/2$ in between the two other peaks.
Depending on its phase whose main contribution is $\epsilon\tau_1$, the
interferences will be destructive or constructive.

Figure \ref{f2} shows the integrated FWM signal as $\tau_1$ is tuned within
a narrow range around  $\tau_1=41\pi/\epsilon=0.23T_{ph}$. At time 
$\tau_1=0.227T_{ph}$ the different contributions in Eq. (\ref{eq6}) add up 
constructively to form a single broad peak at $\tau=T_{ph}-\tau_1/2$. At 
$\tau_1=0.231T_{ph}$, however, the destructive interference splits the peak 
into two bumps separated by a time $\tau\simeq 0.4T_{ph}$. The separation 
exceeds $\tau_1$ because the interference suppresses the signal in the 
whole range of $\tau$ where the peaks $G(\tau,0)$ and $G(\tau+\tau_1,0)$ have a 
significant overlap.

The splitting of the phonon peaks was interpreted in Ref. 
\onlinecite{weg3} as a doubling of frequency due to two-phonon processes
since the peaks were separated by approximately $T_{ph}/2$ 
for the values of $\tau_1$ investigated.  Within the two-level model, however, 
the additional peak that appears due to destructive interference, clearly 
moves with varying $\tau_1$ and cannot be interpreted as a frequency doubling. 
Figure \ref{f3} shows the integrated FWM signal for
various $\tau_1$ tuned as to produce well separated doublet peaks, $P^-$ 
and $P^+$, for time delays close to $\epsilon\tau_1=31,41,51$ and $61\pi$. The 
$P^-$ peak follows the dispersion $T_{ph}-\tau_1$ as illustrated by the inset 
which shows the positions of the maxima $\tau_{max}$ as a function of $\tau_1$ 
for coupling strengths $g=0.8$ and $1.5$. While for $g=0.8$ the rather broad 
peaks are splitted by more than $\tau_1$, for $g=1.5$ their positions
approach the expected values of $\tau_{max}=T_{ph}-\tau_1$ and $T_{ph}$ 
because the peaks have a reduced overlap.

The FWM signal for fixed $\tau$ but varying $\tau_1$ shows bursts 
at multiples of the phonon period which are a clear signature of the strong 
coupling regime. Figure \ref{f4} shows the intensity of the first phonon 
peak $I(\tau=T_{ph},\tau_1)$ rapidly oscillating with the interband frequency 
$\epsilon$.  The envelopes $I_{\pm}(\tau_1)$ of the oscillations shown as 
thick lines, vary exponentially with $x=\sin^2(\Omega\tau_1/2)$ for
large $g$: 
\begin{eqnarray}
I_{\pm}(\tau_1)\propto 1+\exp\left(-2\Gamma\tau_1-2g^2(1+4x-\sqrt{1+8x})\right)
\pm\nonumber\\
2\exp\left(-\Gamma\tau_1-2g^2(1+2x-(1+2x+\sqrt{1+8x})/2)^{1/2}\right)
\end{eqnarray}
For large $g$ the envelopes exhibit bursts at multiples of the phonon period.
The bursts are a non-perturbative effect involving all the
many harmonics that contribute to the FWM signal. For an arbitrary time
delay $\tau_1$ the harmonics which are out of phase destroy the phase
coherence induced by the pulses, and the oscillations are suppressed 
exponentially rapidly with $\tau_1$. For resonant times $\tau_1$ however,
the different harmonics add up coherently and the phase sensitivity is
recovered.

In conclusion we have studied the FWM response of a two-level system coupled 
to a single phonon mode focusing on the strong coupling regime. Analytical 
calculations of the integrated FWM signal provided new insight into the 
interplay between coherent control and dephasing through scattering processes:
(i) The FWM signal has sharp peaks at multiples of
the phonon period whose width decrease as $1/\sqrt{g}$ with increasing coupling.
(ii) Two phase-locked pulses delayed by $\tau_1$ give rise to two sets of
peaks which add up coherently into a single set of bump at 
$nT_{ph}-\tau_1/2$ for $\epsilon\tau_1$ a multiple of $\pi$. When the two 
signals are out of phase, however, the peaks split into doublets 
at approximately $nT_{ph}-\tau_1$ and $nT_{ph}$. This picture contrasts with 
the interpretation of the coherent control experiments in terms of
a frequency doubling. (iii) The FWM signal for fixed $\tau$ exhibits bursts 
of coherence at resonant values of $\tau_1=nT_{ph}$, which are a clear signature
of the non-perturbative regime and might be observed in experiments by 
measuring the intensity of the first phonon resonance as a function of $\tau_1$.
\acknowledgments
This work was supported by the Max-Planck Society.

\begin{references}
\bibitem{com} The corresponding nomenclature for time delays in Ref. 1 is
$\tau=t_{21}$ and $\tau_1=t_{11}$.
\end{references}
\end{document}